\documentclass[12pt,letterpaper]{article}

\usepackage{amsmath,latexsym,amssymb,color,hyperref}
\newcommand{\be}{\begin{equation}}
\newcommand{\ee}{\end{equation}}
\newcommand{\bea}{\begin{eqnarray}}
\newcommand{\ena}{\end{eqnarray}}
\newcommand{\no}{\noindent}

\renewcommand\o{\omega}

\newcommand\m{\ensuremath{\mu}}

\newcommand\n{\ensuremath{\nu}}

\newcommand{\bm}{{\bf m}}

\newcommand{\bC}{{\bf C}}

\newcommand{\bP}{{\bf P}}
\newcommand\tr{\text{tr}}
\newcommand\diag{\text{diag}}

\newcommand{\de}{\partial}

\newcommand{\sch}{Schwarzschild}
\newcommand{\ba}{\begin{eqnarray}}
\newcommand{\ea}{\end{eqnarray}}

\makeatletter
\def\ps@mine{%
    \def\@oddfoot{\hfil\thepage\hfil}\let\@evenfoot\@oddfoot
    \let\@oddhead\@evenhead%
    \let\@mkboth\@gobbletwo
    \let\sectionmark\@gobble
    \let\subsectionmark\@gobble
    }
\renewcommand\section{\@startsection {section}{1}{\z@}%
                                   {-3.5ex \@plus -1ex \@minus -.2ex}%
                                   {2ex \@plus.2ex}%
                                   {\normalfont\large\sffamily\bfseries}}
\renewcommand\subsection{\@startsection {subsection}{1}{\z@}%
                                   {-3.5ex \@plus -1ex \@minus -.2ex}%
                                   {2ex \@plus.2ex}%
                                   {\normalfont\sffamily\bfseries}}
\makeatother

\begin{document}

\def\FILL{\hfill\hfill\hfill}

\title{\sffamily\bfseries 
  Spherically Symmetric Solutions in\FILL\\
 Ghost-Free  Massive Gravity\FILL\\[3ex]
  \normalsize
  D. Comelli$^a$, M. Crisostomi$^{b,c}$, F. Nesti$^{b,d}$ and   L. Pilo$^{b,c}$\FILL\\[2ex]
  \it\small
  $^a$INFN, Sezione di Ferrara,  I-35131 Ferrara, Italy\FILL\\
  $^b$Dipartimento di Fisica, Universit\`a di L'Aquila,  I-67010 L'Aquila, Italy\FILL\\
  $^c$ INFN, Laboratori Nazionali del Gran Sasso, I-67010 Assergi, Italy\FILL\\
  $^d$ ICTP, I-34100 Trieste, Italy\FILL\\[2ex]
  {\tt comelli@fe.infn.it}, {\tt marco.crisostomi@aquila.infn.it},\FILL\\
  {\tt fabrizio.nesti@aquila.infn.it}, {\tt luigi.pilo@aquila.infn.it}\FILL\\[-6ex]
}

\date{\small \today}
\maketitle

\thispagestyle{empty}

\vspace*{-6ex}

\def\abstractname{\sc Abstract}
\begin{abstract}
  \noindent
  Recently, a class of theories of massive gravity has been shown to
  be ghost-free.  We study the spherically symmetric solutions in the
  bigravity formulation of such theories.  In general, the solutions
  admit both a Lorentz invariant and a Lorentz breaking asymptotically
  flat behaviour and also fall in two branches. In the first branch,
  all solutions can be found analitycally and are Schwarzschild-like,
  with no modification as is found for other classes of theories.  In
  the second branch, exact solutions are hard to find, and relying on
  perturbation theory, Yukawa-like modifications of the static
  potential are found.  The general structure of the solutions
  suggests that the bigravity formulation of massive gravity is
  crucial and more than a tool.
\end{abstract}

\bigskip

\section{Introduction}

Recently, there has been a renewed interest in the search of a
modified theory of gravity at large distances through a massive
deformation of GR  
(see for a recent review ~\cite{Hint}), trying to extend at the
nonlinear level~\cite{Gabadadze:2010}
the seminal work of Fierz and Pauli (FP)~\cite{Fierz:1939ix}.
FP is defined at linearized level and is
plagued by a number of diseases. In particular, the modification of
the Newtonian potential is not continuous when the mass $m^2$
vanishes, giving a large correction (25\%) to the light deflection
from the sun that is experimentally excluded~\cite{DIS}. A possible
way to circumvent the physical consequences of the discontinuity was
proposed in~\cite{Vainshtein}; the idea is that the linearized
approximation breaks down near a massive object like the sun and an improved perturbative
expansion must be used that leads to a continuous zero mass limit.
In addition, FP is problematic as a
n effective theory. Regarding FP as
a gauge theory where the gauge symmetry is broken by a explicit mass
term $m$, one would expect a cutoff $\Lambda_2 \sim m g^{-1} = (m
M_{pl})^{1/2}$, however the real cutoff is $\Lambda_5 = (m^4
M_{pl})^{1/5}$ or $\Lambda_3 = (m^2 M_{pl})^{1/3}$, much
lower than $\Lambda_2$ \cite{AGS}. A would-be Goldstone mode is responsible for
the extreme $UV$ sensitivity of the FP theory, that becomes totally
unreliable in the absence of proper UV completion.  
Recently it was
shown that there exists a non linear completion of the FP theory~\cite{Gabadadze:2011}
that is free of ghosts, avoiding the presence of the Boulware-Deser
instability~\cite{BD}.
Then the propagation of only five degrees of freedom and the absence of
instabilities was generalized in~\cite{GF}; this was shown also in the Stuckelberg language in~\cite{deRham:2011qq}.

Quite naturally massive gravity leads to bigravity.  Indeed, any
massive deformation is obtained by adding to the Einstein-Hilbert
action a non-derivative self-coupling of the metric $g$, requires the
introduction of an additional metric $\tilde g$. This auxiliary metric
may be a fixed external field, or be a dynamical one.  When $\tilde g$
is non dynamical we are dealing with \ae ther-like theories; on the
other hand if it is dynamical we enter in the realm of
bigravity~\cite{DAM1} that was originally introduced by Isham, Salam
and Strathdee~\cite{Isham}.  One of the benefits is that such a
theories are automatically diff invariant.  The need for a second
dynamical metric also follows from rather general grounds. Indeed, it
was shown in~\cite{DeJac} that in the case of non singular static
spherically symmetric geometry with the additional property that 
the two metrics are diagonal in the same coordinate patch, a Killing
horizon for $g$ must also be a Killing horizon for $\tilde g$.  Thus,
it seems that in order that the Vaishtein mechanism is effective and
GR is recovered in the near horizon region of a black hole, $\tilde g$
has to be dynamical. The bigravity setup naturally leads one to
explore the possible Lorentz breaking in the gravitational sector, due
to different coexisting backgrounds.

After a brief discussion in section~\ref{bi} on how massive gravity
can be casted in a class of bigravity theories, in section~\ref{flat}
we present a detailed analysis of both the flat Lorentz invariant and
Lorentz breaking phases for the ghost-free potential recently found.
In section~\ref{sec:spherically} we study the spherically symmetric
type I solutions where the second metric is non diagonal, and type II
solutions with codiagonal metrics. In section~\ref{sec5} we compare
the bigravity solutions to those found in Stuckelberg approach.
Section~\ref{con} contains our conclusions.

\section{Massive Gravity as Bigravity}
\label{bi}
Any modification of GR turning a massless graviton into a massive one
calls for additional DoF. An elegant way to provide them is to work
with the extra tensor $\tilde g_{\mu \nu}$. When coupled to the
standard metric $g_{\mu \nu}$, it allows to build non-trivial
diff-invariant operators that lead to mass terms, when expanded around
a background.  Consider the action
\be
S=
\int d^4 x  \sqrt{\tilde g} \,   \kappa   \, M_{pl}^2\; \tilde {\cal R}+\sqrt{g} \left[ 
M_{pl}^2 \left( {\cal R}
-2  m^2   \, V \right)  + L_{\text{matt}} \right] ,
\label{act} 
\ee
where ${\cal R}$ and $\tilde {\cal R}$ are the corresponding Ricci
scalars, and the interaction potential $V$ is a scalar function of the
tensor $X^\mu_\nu = {g}^{\mu \alpha} {\tilde g}_{\alpha \nu}$.  Matter
is minimally coupled to $g$ and it is described by $L_{matt}$. The
constant $\kappa$ controls the relative  strength of
gravitational interactions in the two sectors, while $m$ sets the
scale of the graviton mass. The action (\ref{act}) brings us into the
realm of bigravity theories, whose study started in the 1960s (see
\cite{DAM1} for early references).  An action of the form (\ref{act})
can be also viewed as the effective theories for the low lying
Kaluza-Klein modes in brane world models~\cite{DAM1}. An additional
matter sector can be minimally coupled to $\tilde g$, see for
instance~\cite{us}.  The massive deformation is encoded in the
nonderivative 
coupling between $g_{\mu\nu}$ and the extra tensor field
$\tilde g_{\mu \nu}$. Clearly the action is invariant under
diffeomorphisms, which transform the two fields in the same way
(diagonal diffs).\footnote{The gauge symmetry may be further enlarged
  to the full set of $\text{Diff}_1 \times \text{Diff}_2$ by
  introducing a suitable set of Stuckelberg fields~\cite{AGS}.}
Taking the limit $\kappa \to \infty$ the second metric decouples, and
gets effectively frozen to a fixed background value and diffs are
effectively broken.  Depending on the background value of $\tilde
g_{\mu \nu}$ one can explore both the Lorentz-invariant (LI) and the
Lorentz-breaking (LB) phases of massive gravity.
The role played by $\tilde g_{\mu \nu}$ is very similar to the Higgs
field, its dynamical part restores gauge invariance and its background
value determines the realization of the residual symmetries.

The modified Einstein equations can be written as\footnote{When not
specified, indices of tensors related with $g$($\tilde g$) are raised/lowered with
$g(\tilde g)$.}
\begin{gather}
\label{eqm1}
\,{E}^\mu_\nu  +  Q{}^\m_\n = \frac{1}{2\;M_{pl}^2 }\, {T}^\mu_\nu   \\
\label{eqm2}
\kappa  \, {\tilde E}^\mu_\nu  +  \tilde Q{}^\m_\n = 0 \; ;
\end{gather}
where we have defined $Q$ and $\tilde Q$ as effective energy-momentum
tensors induced by the interaction term. The only invariant tensor
that can be written without derivatives out of $g$ and $\tilde g$ is
$X^\mu_\nu = g^{\mu \alpha} {\tilde g}_{\alpha \nu}$ \cite{DAM1}. The
ghost free potential~\cite{Gabadadze:2011} $V$ is a special scalar function of
$Y^\mu_\nu=(\sqrt{X})^\mu_\nu$ given by
%~\cite{BCNP}
%
\be
\label{eq:genpot}
\qquad V=\sum_{n=0}^4 \, a_n\, V_n \,,\qquad n=0\ldots4
\vspace*{-2ex}
\ee 
\vskip .3cm
\no
where $V_n$ are the symmetric polynomials of $Y$
%
%\vspace*{-1ex}
\be
\begin{split}
&V_0=1\,\qquad 
V_1=\tau_1\,,\qquad
V_2=\tau_1^2-\tau_2\,,\qquad
V_3=\tau_1^3-3\,\tau_1\,\tau_2+2\,\tau_3\,,\\[1ex]
&V_4=\tau_1^4-6\,\tau_1^2\,\tau_2+8\,\tau_1\,\tau_3+3\,\tau_2^2-6\, \tau_4\,,
\end{split}
\ee 
with $\tau_n=\tr(Y^n)$. As a result we have
\bea
\label{eq:q1}
 Q{}_\nu^\mu &=&  { m^2}\, \left[ \;  V\; \delta^\mu_\nu \,  - \,  (V'\;Y)^\mu_\nu  \right]\\[1ex]
\label{eq:q2}
 \tilde Q{}_\nu^\mu &=&  m^2\, q^{-1/2} \, \; (V'\;Y)^\mu_\nu ,
\ena
where  $(V^\prime)^\mu_\nu = \de V / \de Y_\mu^\nu$ and  $q =\det
X=\det(\tilde g)/\det(g)$.

\section{Flat Solutions}
\label{flat}
It is interesting to study the structure of both the Lorentz invariant
(LI) and Lorentz breaking (LB) phases, starting from the (bi)flat
solutions, which will be the benchmark for the asymptotic behaviour in
the general case.  Generically, equations (\ref{eqm1}) and
(\ref{eqm2}) admit the solutions~\cite{us,PRLus}
\be
g=\eta=\diag(-1,1,1,1) \, , \qquad  
\tilde g= \tilde\eta= \o^2 \, \diag(- c^2,1,1,1) \,,
\label{bkg}
\ee
where $c$ and $\omega$ are parameters to be determined by $V$, when
one imposes $Q=\tilde Q=0$.  Let us discuss the LI ($c=1$) and LB
($c\neq1$) cases, where these conditions give respectively two and
three independent equations.

%

%\bigskip \centerline{\bf LI Phase} \medskip

\subsection{LI Phase}
\no For the flat LI background ($c=1$) the conditions are as follows:
\be\label{fli}
\begin{split}
&a_0= 12\, \omega ^2 \,\left(6 \,a_4 \omega ^2+4 \,a_3\,
   \omega +a_2\right) , \\ 
&a_1= -6 \,\omega \, \left(4\, a_4 \omega^2 +3 \,a_3\omega+a_2\right) .
\end{split}
\ee
Having two equations for a single parameter $\omega$, this means that
in the LI phase one fine tuning is needed.  This corresponds to the
standard tuning of the cosmological constant.

Since the chosen potential (\ref{eq:genpot}) is ghost-free, the
quadratic mass term is automatically of the FP form
\be
\begin{split}
&L_{m}=- m_g^2 \,
\left( h^t_{\mu \alpha} \bP h_{\nu \beta} - \, h^t_{\mu \nu} \bP h_{\alpha \beta} \right)\, 
\eta^{\mu \alpha} \eta^{\nu \beta} \, , \\[3ex] 
&\bP = \begin{pmatrix} \,1 & -1 \\ -1\, & 1  \end{pmatrix} \, ;\\
\end{split}
\ee
where we introduced the 2-component column vector: $h_{\m\n} \!=\!
(h_{\m\n},\tilde h_{\m\n})^t$ containing the two metric perturbations
$g_{\mu \nu} = \eta_{\mu \nu} + h_{\mu \nu}$ and $\tilde g_{\mu \nu} =
\tilde \eta_{\mu \nu} +\o^2\; \tilde h_{\mu \nu}$. The mass parameter
is
\be
\label{mgLI}
m_{g}^2=-m^2 \,\omega ^2\, \left(12 \, a_4 \, \omega^2
   +a_3\omega+a_2\right) .
\ee
At the linearized level, a massless spin two and spin two with mass
$m_g$ are the propagating modes.

\subsection{LB Phase}
%\bigskip \centerline{\bf LB Phase} \medskip

\no When $c \neq 1$ Lorentz symmetry is broken by the VEV of $\tilde
g$.  The conditions are three:
\be
\begin{split}
 a_0&= -24 \,\omega ^3 \left(3\, a_4 \,\omega
   +a_3\right) \, , \qquad  a_1=6\, \omega ^2 \left(8\, a_4 \,\omega +3
  \, a_3\right) \, , \\
a_2 & = -6\, \omega  \left(2\, a_4 \,\omega+a_3\right) \, .
\end{split}
\label{flb}
\ee
Among these, one determines $\omega$ in terms of the $a_i$, the
remaning two are fine tunings, equivalent to setting to zero the
effective cosmological constants for each of the two metrics.
Moreover, we note immediately that $c$ is not determined by the above
equations, and is thus a free parameter.  The situation is to be
compared to bigravity with generic potentials \cite{PRLus}, where two
equations determine $\omega$ and $c$, and only one fine tuning is
required for biflat solutions. Hence, this is a peculiarity of the
potentials (\ref{eq:genpot}) considered. The fact that $c\neq1$ is not
determined, suggests that the potential has some flat directions for
the metric fluctuations.  This peculiar behaviour is due to the fact
that for the ghost free potential the lapse N, related to $g_{tt}$ in
ADM formalism, is a Lagrange multiplier. This is confirmed already at
quadratic level.

The expansion of the potential at quadratic order gives a generic LB
mass term of the form
\be
\label{mass}
%\begin{split}
L_{mass} =
\frac14\Big( h_{00}^t \bm_0 h_{00}
     +2 h_{0i}^t \bm_1 h_{0i}  -  h_{ij}^t \bm_2 h_{ij} 
%\nonumber\\ 
 % \!&&\!{} \qquad 
+ h_{ii}^t \bm_3 h_{ii} 
     - 2  h_{ii}^t \bm_4 h_{00} \Big)\,.
\ee
%\end{eqnarray}
%
While for a generic potential we have the following
matrix structure~\cite{PRLus}
\be
\begin{array}[c]{lcl}
\bm_0 &=& \lambda_0 \, \bC^{-2} \bP \bC^{-2}   \\[.8ex]
\bm_1&=&0  \\[.8ex]
\bm_{2,3} \!&=&  \lambda_{2,3}\, \bP \\[.8ex]
\bm_4 &=& \lambda_4 \,\bC^{-2}\bP  
\end{array}
%\hspace{-4.5cm}
\begin{array}[c]{lcl}
%\begin{split}
&\bP = \begin{pmatrix} \,1 & -1 \\ -1\, & 1  \end{pmatrix} \\[3ex]
&{\bf C} = \diag (1, c) \, ,\\
%\end{split}
\end{array}
\label{eq:masses}
\ee
for the potential (\ref{eq:genpot}), the masses turn out to be
\be
\label{lambdas}
\begin{split}
&\lambda_0 =\lambda_4 = 0 \\
&  \lambda_2 =\lambda_3=\frac{3}{2}\,  m^2 \, (1-c) \,\omega ^3 \,\left(4\, a_4 \,\omega +a_3\right).
\end{split}
\ee 
The mass of the spatial transverse traceless propagating mode (2 DoF)
is proportional to
\be
\label{mgLB} 
m_{g\,LB}^2=m^2 \, \frac{3}{2} \, (1-c) \, \omega ^3 \, \left(4 \, a_4
  \,\omega +a_3\right) \, .
\ee
For $c < 1$ its positivity requires $(4\, a_4 \,\omega +a_3)>0$
while for $c > 1$, it requires $(4\, a_4 \,\omega +a_3)<0$.  Note that
in the limit $c \to 1$, the LB phase intersects the LI phase but the
graviton is massless, i.e.  $m_{g\,LB}\to0$.

Because $\det(\bP)$=0, together with the massive tensor mode there is
always a massless one in the spectrum of metric perturbations.  The
corresponding phenomenology is quite rich, and was analyzed
in~\cite{PRLus}.  The linearized theory can be interpreted as a
diff-invariant realization of massive gravity, free of ghosts and
phenomenologically viable (no vDVZ discontinuity is
present).\footnote{Theories of Lorentz-breaking massive gravity were
  analyzed also in~\cite{RUB,DUB}.}  The only propagating degrees of
freedom at linearized level are the spatial transverse traceless
tensor modes (2 polarizations for each metric) physically representing
a massless and a massive graviton (gravitational waves) oscillating
one in the other and with different speeds, resulting in a nontrivial
dispersion relation.  The possibly superluminal speed $c^2$ in the
second gravitational sector does not lead to causality violations,
because the new metric has the character of '\ae ther'. The physical
consequence is that gravitational wave experiments become
frame-dependent.

Finally, from the analysis of~\cite{PRLus} (see table I there), we see
that since $\lambda_0=\lambda_4=0$, at linearized level a degree of
freedom is not determined, and an extra gauge mode corresponding to a
shift in $h_{00}-\tilde{h}_{00}$ appears. This is an artifact of the
linearized approximation, because we know~\cite{Gabadadze:2011} that at nonlinear
level no additional gauge invariance is preserved (beyond the four
diffs). Therefore we expect this mode to be determined at the
nonlinear order.

\section{Spherically Symmetric Solutions in Vacuum}
\label{sec:spherically}

In GR, the form of the exterior solution for a spherically symmetric
self-gravitating body is dictated by the Birkoff theorem to be
Schwarzschild.  Since in our case the vacuum Einstein equations are
modified by the presence of the tensors $Q$, $\tilde Q$ we expect that
spherical solutions may deviate from Schwarzschild.

In a spherically coordinate system $(t,r,\theta, \varphi)$, the form of $g$
and $\tilde g$ is
\be
\begin{split}
\label{sm1}
ds^2 &= - J(r) \, dt^2 + K(r) \, dr^2 + r^2 \, d \Omega^2\,,\\[.7ex]
d\tilde s^2 &= - C(r) \, dt^2 + A(r) \, dr^2 + 2\; D(r) \, dt\, dr +  B(r) \, d \Omega^2 \, .
%\label{sm2}
\end{split}
\ee
Because of $D$, in general we cannot bring both metrics in a diagonal form
with a coordinate transformation.

Solutions fall into two classes \cite{Isham,chelaflores}: type I for
$D\neq 0$ and type II for $D= 0$.  Since $E_{\mu}^\nu$ is diagonal by
the choice of the first metric, then also $(V'Y)^\mu_\nu$ must be
diagonal because of (\ref{eqm1}). The only possible source of
off-diagonal term in RHS of (\ref{eqm2}) would be $(V'Y)^\mu_\nu$, and
as a result also $\tilde E_{\mu}^\nu$ must be diagonal.  In general,
the off-diagonal components of $Q$ and $\tilde Q$ are of the
form 
\be Q_{t}^r=Q_{r}^t \propto \;D(r)\; \left( 4 a_2 r
  B^{1/2}+6 a_3 B+a_1 r^2 \right) \propto 0 \, .  
\ee 
Thus, we have two options: either $D(r)=0$ or $B(r)=\o^2\,r^2$ with
\be
6 \,\omega ^2  \, a_3 +4\,\omega \, a_2  +a_1 =0 \, .
\label{oeq}
\ee
Comparing this condition with the conditions to have the flat
background solution (\ref{fli}) or (\ref{flb}), we realize that in the
LB scenario such a condition is automatically realized, while in the
LI one an extra constraint is required:\footnote{Note that this
  corresponds to a massless LI graviton, $m_g=0$ see
  eq. (\ref{mgLI}).}
\be
 12 \,a_4\, \omega^2+6\,a_3\,\omega +a_2=0\, .
\label{extra}
\ee
Thus, for $D\neq 0$, the conditions required in the LB and LI phases
are the same.

\subsection{Type I Solutions}
\label{tone}

For $D(r)\neq 0$ the solution can be found  analytically and reads
\be
\begin{split}
J&= 1-\frac{2 \, m_1}{r} + \Lambda_1 \;r^2 \, , \quad
K(r)=\frac{1}{J}  ,  \quad
B=\omega^2r^2\, , \quad
D^2+A\, C=c^2\omega^4\vspace*{-3ex}\\[1ex]
C&= c^2\omega^2 \left(1-\frac{ 2 \kappa^{-1}  m_2}{r} + \Lambda_2 \;r^2 \right)  ,
\quad A= \frac{(c^2+1) \;\omega ^2 \;J- C}{J^2} \, ,
\end{split}
\label{one}
\ee
where $m_{1,2}$ and $c$ are integration constants, 
 $\omega$ satisfies eq.~(\ref{oeq}),
and
\bea
\label{L1}
\Lambda_1 &=& -\frac{1}{3} m^2 \left(a_0-12\, a_3\, \omega^3
   -6\,a_2\,\omega^2\right) ,\\
\label{L2}
 \Lambda_2 &=&-\frac{2\, m^2}{3\,
   \kappa } \left(12\, a_4\, \omega ^2+6\, a_3\, \omega +a_2\right) .
\ena
As a result, the geometry for both metrics is dS(AdS) \sch.  It is
remarkable that the deviation from \sch\ which is present in those
exact solutions for different (quite similar) potentials~\cite{us}, in
the form of a nonanalytic term $r^\gamma$, is totally absent for this
choice of potentials.

Of course, if we require the solution to be asymptotically flat, then
$\Lambda_{1,2}=0$ and the three conditions (\ref{flb}) must be
satisfied, for both the LI and LB cases.\footnote{Eq.~(\ref{oeq}) and eqs.~(\ref{fli}) together, are equivalent to eqs.~(\ref{flb}).}
While in the LB phase this leaves space for a massive graviton
($m_{g}=m_{g\,LB}\neq 0$), in the LI phase eq.~ (\ref{extra}) forces the
mass of the spin 2 mode to be zero ($m_{g}= 0$).\footnote{Note: the
  solution~(\ref{one}) is valid only if $a_3+4\, \o\,a_4\neq 0$. In
  case this quantity vanishes, the solution (\ref{one}) is still valid
  except that $A$ disappears from the equations of motion and remains
  undetermined. Also, in this case at linearized level all LB masses
  (\ref{lambdas}) vanish.}

A remarkable property of the type I solutions is that independently on
the potential $JK=1$; as a result no vDVZ discontinuity is present.
The constants $m_1$ and $m_2$ are related to the total gravitational
mass $m_{tot}$ of the system. In fact a study of the total energy (in
the asymptotically flat case) showed~\cite{energy-us} that $m_{tot} =
m_1+ m_2$, with interesting cancellation of the mass screening
mechanism in the case of realistic star solutions.

The key point in deriving the solution (\ref{one}) is that once $B=
\omega^2 \;r^2$, then $Q$ and $\tilde Q$ behave exactly like the
energy momentum of a cosmological constant; this fact is also
interesting in view of cosmological applications.

From the linearized LI phase one expects that a combination of a
massless and massive spin 2 modes should mediate gravitational
interactions, however surprisingly for type I solution only the
massless field is important as it is evident from the $1/r$ behaviour
of the solution (\ref{one}).  The absence of a Yukawa exponential
suppression of the static gravitational potential can be explained by
the fact that when (\ref{oeq}) is satisfied, the FP mass vanishes, see
eq.~(\ref{mgLI}).  As far the LB phase is concerned, the values of the
masses are such that all scalar modes mediate $1/r$ instantaneous
interactions~\cite{PRLus}\footnote{From eqs.(\ref{lambdas}) one can
  see that the paramater $\lambda_\mu$ defined in \cite{PRLus} which
  controls the deviation from $1/r$ at the linearized level,
  vanishes.}  though there are massive tensor modes that propagate.
This is the main difference between the LI and LB phase.

\subsection{Type II Solutions}
\label{ttwo}

For $D=0$ the equations of motion are in general very difficult to
solve analytically. In general even in this case we can have LI or LB
flat asymptotics.

In this section we focus on the LI case, the discussion on the LB case
can be found in appendix~\ref{twoLB}.

A class of exact solutions can be found if one takes as an ansatz the
form of $B$ that was obtained for the type I solutions, namely $B =
\omega^2 \, r^2$ with $\o$ satisfying (\ref{oeq})(see
also~\cite{Nieuwenhuizen:2011sq}).  Then one gets that the metrics are
conformally related
\be
\tilde g_{\mu \nu} = \omega^2 \, g_{\mu \nu} \, , 
\ee
with
\be
%\begin{split}
J= 1-\frac{2 \, m}{r} + \bar \Lambda r^2 \, , \qquad KJ=1 \, , \qquad
B=\omega^2r^2\, , \;
% & \bar \Lambda = \frac{2}{3 \, \kappa}  \left[6 \omega  \left(2 a_4 \omega%     +a_3\right)+a_2\right]
% \end{split}
\label{twop}
\ee
where $m$ is an integration constant and the value of $\bar \Lambda$
coincides with $\Lambda_1$ of type I solutions, eq.~(\ref{L1}). Again,
we have a dS(AdS) \sch\ solution but, differently from type I
solutions, both metrics are simultaneously diagonal with a single
integration constant.

If no ansatz on $B$ is given, one is unable to solve the highly
nonlinear set of equations. Nevertheless a weak field expansion is
clearly viable: using the consistent background $g_{\mu \nu} =
\eta_{\mu \nu}$, $\tilde g_{\mu \nu} = \omega^2 \, \eta_{\mu \nu}$,
and setting
% \footnote{We have implemented
% the consistency of the background by 
% choosing $a_0$ and $a_2$ in a such way that ${Q}_{g=\eta \, , \tilde
%   g =
%   \omega \eta}={\tilde Q}_{g=\eta \, , \tilde g =
%   \omega \eta}=0$}
%
\be
\begin{split}
&J = 1 + \epsilon \, \delta J \,, \quad \delta K =  1 + \epsilon \, \delta K 
\, , \quad  C=\omega^2(1 + \epsilon \, \delta C) \,, \\
&A=\omega^2(1 + \epsilon
\, \delta A) \, , \quad B =\omega^2 (r^2 + \epsilon  \, \delta B)\;,
\end{split}
\ee
from the equations of motion at order $\epsilon$ one finds
\be
\begin{split}
\delta K&= -\frac{\, \, m_2 \, e^{-\frac{r}{\lambda_g}}
  \; (r+\lambda_g )}{r\, \lambda_g }+\frac{2\,
   m_1}{r}\\
   \delta  J&= \frac{2\, m_2 \, 
   e^{-\frac{r}{\lambda_g }}}{r}-\frac{2\,
   m_1}{r} \\
   \delta B&=  m_2  \, e^{-r/\lambda_g}  \, \frac {\left(\kappa  \omega ^2+1\right)
 \left(\lambda_g^2+r^2+\lambda_g  r \right)}{\kappa  \, \omega ^2 \, r }\\
   \delta C&= -\frac{2 m_1}{r} -\frac{2 \,  m_2 \, 
   e^{-r/\lambda_g}}{\kappa \,  \omega ^2\,  r}\\
   \delta A&= \frac{2 m_1}{r} -     \, m_2 \, e^{-r/\lambda_g} \frac{ (\lambda_g+r) \left[2 \lambda_g^2
   \left(\kappa  \, \omega ^2+1\right)+\kappa\,   \omega^2  \, r^2
 \right]}{\kappa  \, \omega ^2\, \lambda_g \, r^3 }
\end{split}
\label{pert}
\ee
where $\lambda_g^{-1} =m_g \, \sqrt{ 2(\omega^{-2} \kappa^{-1}+1) } $.

The solution clearly shows the vDVZ discontinuity. In fact, in the
limit $m_g\to0$ ($\lambda_g \to \infty$) one has
\be
\delta J+\delta K=
\frac{ m_2\, e^{-\frac{r}{\lambda_g }} \, (\lambda_g  - r)}{r
  \lambda_g}  \  \to \  \frac{m_2}{r},
\ee
which does not vanish as it does in GR.  Actually the weak field
expansion is not even well defined in the $m \to 0$ limit; indeed in
this limit the perturbations $\delta B$ and $\delta A$ diverge.
Notice however that though $\tilde g_{\mu \nu}$ is singular in the
limit $\lambda_g \to \infty$, the associated Riemann tensor is well
defined for any $\lambda_g$. This suggests that the singular behaviour
is due to the choice of coordinates rather than to a real singularity.
We leave the detailed study of this problem, that is related to the
Vainshtein mechanism, for a future work~\cite{usfut}. The Vainshtein
mechanism in the Stuckelberg approach for the ghost free potential was
studied in~\cite{Koyama}, for different potentials see~\cite{others}.

\section{Comparison with the Stuckelberg approach}
\label{sec5}

In this section we compare our previous solutions with the ones
obtained with a frozen auxiliary metric. Our results partially
coincide with \cite{Koyama}.

Massive gravity can be also formulated taking the second metric as an
absolute flat metric, and introducing a suitable set of
``Stuckelberg'' fields to recover diff invariance. Generically, a flat
metric can be written as
\be
\tilde g_{\mu \nu} = E^a_\mu E^b_\nu \eta_{ab} \, , \qquad E^a_\mu =
\de_\mu \Phi^a \;. 
\label{frozen}
\ee
The four Stuckelberg fields $\Phi^a$ are used to parametrize the ``flat''
 vielbein and physically represents the global ``flat'' coordinates in
 which the metric $\tilde g_{\mu \nu}$ is flat. In this formulation of
massive gravity the action is
\be
S=
\int d^4 x  \sqrt{g} \left[ 
M_{pl}^2 \left( {\cal R}
-2  m^2   \, V \right)  + L_{\text{matt}} \right] ,
\label{actfr} 
\ee
The potential is the same of (\ref{eq:genpot}) and the equation of
motion are just the ones of~(\ref{eqm1}). 

In the spherically symmetric case one can choose coordinates such
that $\Phi^a= \delta^a_\mu$ and 
\be
\begin{split}
\label{sm1}
ds^2 &= - F(r) \, dt^2 + W(r) \, dr^2 + 2 Z(r) \, dt\, dr +  P(r) \, d \Omega^2 \, ,
\\
d\tilde s^2 &= -  \, dt^2 + \, dr^2 + r^2 \, d \Omega^2 \, .
\end{split}
\ee
Such a choice is sometimes called unitary gauge. 

\paragraph{Type I solutions.} For $Z\neq 0$ we have the exact solutions
\be
\begin{split}
&P= r^2 \alpha^2  \qquad  (\alpha \text{ positive root of } \; a_1
\alpha^2+4 a_2 \alpha +6 a_3 =0 )\, ,\\[1ex]
& F= f_0^2 - \frac{2 M}{r} + r^2 \, \Lambda \, , \qquad \Lambda= -
\frac{ m^2 \, f_0^2 \left( \alpha ^3 \, a_0-6 a_2 \, \alpha -12 a_3\right)}{3
 \alpha } \, ,\\[1ex]
&F W+ Z^2 =\alpha^2\;
f_0^2 \, , \qquad  W + F  = f_0^2 + \alpha^2 \,,\\
\end{split}
\label{sol1f}
\ee
where $f_0$ and $M$ are integration constants. As usual when $f_0 =
\alpha$ we are in a LI phase. The requirement to have solution of the form
$g_{\mu \nu} =\alpha^2 \eta_{\mu \nu}$ is
\be
a_0 \, \alpha ^3+3 a_1 \alpha ^2+6 \left(\alpha  \, a_2+a_3\right) =0
\, .
\label{ff}
\ee
When $\alpha$ satisfies the algebraic equation given in (\ref{sol1f}),
the condition for  a conformally flat solution corresponds to $\Lambda
=0$.

As a general comment, in the bigravity formulation we showed that all
solutions of type I are of the form (\ref{one}). In the Stuckelberg
formulation the situation is less favorable and an ansatz is required.

\paragraph{Type II solutions.} Here $Z=0$. In this case, again we have
to rely on perturbation theory.  Expanding around Minkowski space
$P=r^2+\delta P$, $F=1+\delta F$, $W=1+\delta W$, we get
\be
\begin{split} \label{frozen}
\delta F &=
   \frac{-2 c_1 e^{-r /\bar \lambda_g
   }}{r} \, ;\\
   \delta W &=-\frac{2 c_1\bar  \lambda _g e^{-\frac{r}{\bar \lambda
         _g}} \left(\bar \lambda
   _g+r\right)}{r^3} \, ; \\
\delta   P &=\frac{c_1 e^{-\frac{r}{\bar \lambda _g}} \left(\bar
    \lambda _g^2+r\bar  \lambda
   _g+r^2\right)}{r} \, ;
\end{split}
\ee
where $c_1$ is an integration constant and $\bar \lambda_g^{-2} =  m^2\left(a_1+4 \, a_2+6 \, a_3\right)$.

\medskip

Let us compare the above type I and II solutions found in the
Stuckelberg approach with the ones in bigravity; this is possible by
taking the limit $\kappa \to \infty$, which freezes the dynamics of
the auxiliary metric.  For type I solutions in bigravity we have 
\be
\begin{split}
& J \to J \, , \qquad K \to K \, , \qquad C \to c^2 \, \o^2 \, ,
 \\ \nonumber
&A \to A_\infty= \o^2 \, J^{-2} \left[J\; (c^2+1) -c^2 \right]  \, , \quad D \to D_\infty=\left[ c^2 \, \o^2  \,(\o^2 -A_\infty )\right]^{1/2}
\end{split}
\ee
In order to compare to the Stuckelberg approach in the unitary gauge,
we need to change coordinates to bring the second metric in a diagonal
Minkowski form. Taking
\be 
dt= dt^\prime/c\,\o + dr\, D/C \, , \qquad \rho^2 = B(r)\, , 
\ee
we have
\be
\begin{split}
& ds^2 = - J_{\text{new}} \,  d {t^\prime}^2 + K_{\text{new}} \, d \rho^2+2 D_{\text{new}} \, dt^\prime d \rho
+ \o^{-2} \, \rho^2 \, d\Omega^2 \, , \\
& d \tilde s^2 = -d {t^\prime}^2 +  d \rho^2 + \rho^2 \, d\Omega^2 \, , 
\end{split}
\ee
where
\be
\begin{split}
& J_{\text{new}} =\frac{J}{c^2\,\o^2} %c^{-2} \o^{-2} - \frac{2 m1}{c^2 \, \o \, \rho} +\Lambda_1 \, \rho^2 \, \o^{-4} \, c^{-2} 
\, , \quad K_{\text{new}} +
 J_{\text{new}} = \frac{c^2+1}{ \o^{2} \, c^{2}} \, ,\quad
 D_{\text{new}}= - \frac{D_\infty J }{c^3 \omega^4}.
\end{split}
\ee
This gives exactly the solution (\ref{sol1f}) with the identifications
$\alpha=\o^{-1}$ and $f_0= c^{-1} \, \o^{-1}$.

For LI type II solutions the story is different.  In bigravity by
taking the limit $\kappa \to \infty$ in the weak field solution, we
find that $\tilde g_{\mu \nu}$ is not flat, as one can see from the
direct computation of the associated Riemann tensor.  In addition, for
finite $\lambda_g$, there is no choice of $m_1$ and $m_2$ for which
$\tilde g_{\mu \nu}$ is flat. However, the simultaneous limit
$\lambda_g \, , \kappa \to \infty$ leads to a zero Riemann tensor for
$\tilde g_{\mu \nu}$ when $m_1 =0$.  Introducing $\tilde T_{\mu \nu}$,
the EMT  for matter minimally coupled to $\tilde g_{\mu \nu}$,
from the above considerations and from the linearized analysis of
\cite{PRLus} it should be clear that $m_1$ is an integration constant
related to the linear combinations of the sources that couples to the
massless graviton, while $m_2$ is associated with the combination of
sources that couples to the massive graviton.  For instance for
$\kappa=\omega=1$, we have maximal mixing and $m_1 \propto T_{00} +
\tilde T_{00}$, $m_2 \propto T_{00} - \tilde T_{00}$.  Summarizing,
for type II solutions, bigravity in the $\kappa \to \infty$ limit and
the Stuckelberg approach give different results. This is rather
important for studying the Vainshtein effect which is captured by the
type II solutions.

\section{Conclusions}
\label{con}

We have studied the spherically symmetric solutions for massive
gravity in the bigravity formulation of the theory. The interaction
potential that we used was recently shown to be ghost free.  The
theory admits both Lorentz Invariant and Lorentz Breaking flat
solutions, which can be used as asymptotic backgrounds.  Remarkably,
with this interaction potential the amount of Lorentz breaking, i.e.\
the relative speed of light in the two backgrounds, appears as a free
parameter, not determined by the interaction potential.

The spherical solutions falls in two separate classes: type I
solutions with $D \neq 0$, where the second metric is non diagonal and
type II solutions where $D=0$ and the two metrics are simultaneously
diagonal in the same coordinate patch.  For what concerns type I
solutions, we found that are always \sch-like and the effect of the
massive deformation is equivalent, on shell, to a cosmological
constant.  This might be interesting for cosmology. Type I solutions
do not show any modification of the static part, in particular the
``Newtonian'' potential has the standard $1/r$ fall-off. This is to be
contrasted with analogous solutions found with other choices of
interaction potentials~\cite{us}, which show a non-analytic
modification with respect to \sch. In the LI case, which requires a
fine tuning, this can be physically understood from the fact that at
the linearized level the graviton mass vanishes.  On the other hand in
the LB phase, while the static potential is Newtonian, the spectrum
contains a massive graviton tensor with 2 DoF~\cite{PRLus}), which does
not get excited in the spherically symmetric configuration. The other
modes do not propagate at linear level, where an accidental gauge
invariance appears, but are expected to propagate at nonlinear level.

Modified gravity effects appear in type II solutions where both
metrics are diagonal.  In this case, according to the interesting work
of ref.~\cite{DeJac} two static, spherically symmetric, nonsingular,
and diagonal metrics in a common coordinate system must have the same
Killing horizon. These results imply that the Vainshtein mechanism in
this case cannot take place in a black hole when the second metric is
frozen to be Minkowski.  This simple fact shows that bigravity is more
than a tool in formulating massive gravity.  Type II exact solutions
are very hard to find. Except for a special class where the two
metrics are conformally related, one has to rely on perturbation
theory. One can show that the standard weak field expansion (that is
equivalent to a derivative expansion) cannot be trusted when the
graviton mass is small.  Moreover, the solution found, in the limit
$\kappa \to \infty$ differs fron the one found in the Stuckelberg
approach.  We leave the detailed study of these solutions for a future
work~\cite{usfut}.

\begin{appendix}

\section{LB Type II Solutions and Perturbativity}
\label{twoLB}

Let us discuss now the LB case and we limit ourself to asymptotically
flat solutions.  We remark first that Type I solutions in the LB case
cannot be derived in a standard perturbative way. In fact, suppose we
try to find $\delta D$ pertubatively, expanding at the leading order
${Q}^r_t$ or ${\tilde Q}^r_t$ we have
\be
{Q}^r_t \propto \delta D \ \left(
4 a_2 \, \omega+6 a_3 \, \omega^2+a_1 \right) \; .
\ee
However in the LB case, $4 a_2 \, \omega+6 a_3 \, \omega^2+a_1=0$, and
as a result $\delta D$ cannot be determined. The problem can be
overcomed by turning to the non-democratic perturbation theory
discussed in~\cite{us}. The idea is that the metric perturbations are
of order $\epsilon$, except for $\delta D$ which is of order
$\epsilon^{1/2}$. This choice is enough to capture the nonperturbative
features of the solutions as shown in~\cite{us}.

Let us now discuss type II solutions.  Defining the graviton mass as
$m_g=m_{g\,LB}$ (see (\ref{mgLB})), the differential equations of the
perturbations do not close; precisely we have:
\be
\begin{split} 
\delta J' &= \frac{2 G m_1}{r^2}+m_g^2\; \frac{\delta B}{r} \, ,
   \\
  \delta K &= \frac{2 G m_1}{r} \, ,
   \\
   \delta A&= \omega ^2\;
   \left[\left( \frac{\delta B}{r}\right)' +\frac{2 \,G\,
   m_1}{r}\right] \, , 
   \\
   \delta {C}' &=- \frac{m_g^2 }{k }\; \frac{\delta B}{r}+\frac{2 \,G\, m_1\,
   c^2\, \omega ^2}{r^2} \, ;
\end{split}
\ee
where the fluctuation $\delta B$ is not determined at leading order in
the weak field expansion.

\end{appendix}

\end{document}